# Oxygen Defect Engineered Magnetism of La$_2$NiMnO$_6$ Thin Films

Jasnamol P. Palakkal, Thorsten Schneider, Lambert Alff

*Institute of Materials Science, Technische Universität Darmstadt, 64287 Darmstadt,*

*Germany*

**Abstract**

The double perovskite La$_2$NiMnO$_6$ (LNMO) exhibits complex magnetism due to the competition of magnetic interactions that are strongly affected by structural and magnetic inhomogeneities. In this work, we study the effect of oxygen annealing on the structure and magnetism of epitaxial thin films grown by pulsed laser deposition. The key observations are that a longer annealing time leads to a reduction of saturation magnetization and an enhancement in the ferromagnetic transition temperature. We explain these results based upon epitaxial strain and oxygen defect engineering. The oxygen enrichment by annealing caused a decrease in the volume of the perovskite lattice. This increased the epitaxial strain of the films that are in-plane locked to the SrTiO$_3$ substrate. The enhanced strain caused a reduction in the saturation magnetization due to randomly distributed anti-site defects. The reduced oxygen defects concentration in the films due to the annealing in oxygen improved the ferromagnetic long-range interaction and caused an increase in the magnetic transition temperature.

## I. INTRODUCTION

The magnetism of double perovskites (DP) is interesting due to its complex nature. The DP La$_2$BMnO$_6$ with *B*=Cr, Fe, Co, Ni, and Cu are reported to exhibit complex magnetic properties due to competing magnetic interactions, which are affected by structural and magnetic inhomogeneities.[1-5] A near room-temperature magnetic transition ($T_C$) with complex intrinsic and extrinsic magnetic behavior makes the DP La$_2$NiMnO$_6$ (LNMO) interesting.[3] Bulk LNMO has a high-temperature magnetic transition ($T_{C1}$) around 250-270 K and a low-temperature transition ($T_{C2}$) around 100-150 K.[3, 6, 7] Previous reports suggest that the $T_{C1}$ is originating from the orthorhombic LNMO phase containing Ni$^{2+}$ and Mn$^{4+}$ cations, and the $T_{C2}$ is originating from the rhombohedral LNMO phase with Ni$^{3+}$ and Mn$^{3+}$ cations.[6] Recently, this material has attracted even more attention due to its oxygen-reduction and oxygen-evolution reactions caused by its favorable electronic structure.[8] LNMO plays a role in different functional properties like magnetic, multiferroic, optical, and electrochemical activities, as summarized in book chapter.[9] The monoclinic high-$T_C$ phase of bulk LNMO was reported to be stabilized by creating La-vacancies and accommodating extra oxygen in the lattice.[10]

LNMO thin films deposited at high temperature and considerable $pO_2$ were reported to possess a high degree of Ni-Mn cationic ordering with ferromagnetism.[11] A recent study reported that a stoichiometric LNMO thin film produced at a high partial pressure of oxygen ($pO_2$ >300 mTorr) did not show any evidence of ferromagnetic ordering, but Ni-deficiency introduced ferromagnetism in LNMO films.[12] A long-range *B*-site ordering was obtained for films grown at a high temperature of 900 °C.[13] In their study, it was found that the LNMO film prefers a monoclinic structure under tensile and a rhombohedral phase under compressive strain.[13] A recent study on strained epitaxial LNMO films showed that the saturation magnetization ($M_S$) decreases with an increase in strain without altering $T_C$.[14] In general, the multiphase nature, off-stoichiometry, oxygen-deficiency, epitaxial strain, etc., are the controlling factors of magnetism in LNMO. The present work deals with a careful study of oxygen defect engineering in LNMO thin films and control of their magnetic behavior.

## II. EXPERIMENTAL DETAILS

The LNMO thin films were deposited on SrTiO$_3$ (STO) (001) substrates of 5x5 mm$^2$ size by using pulsed laser deposition (PLD). A highly dense LNMO target for the PLD deposition was synthesized by spark plasma sintering the respective oxides at 1050 °C with 8.8 kN force. A laser fluence of 1 J/cm$^2$ was used. The deposition was done at 900 °C with $pO_2$ of 500 mTorr. A five-minute reference time was allowed for the samples at 900 °C before and after deposition. The reference sample code is L5. An in-situ post-annealing at deposition conditions was performed for producing further oxygen defect engineered samples. The samples L35, L65, and L95 have undergone the post-annealing treatment for a duration of 35, 65, and 95 minutes respectively. This corresponds to a decrease in oxygen vacancy concentration with increasing annealing time. X-ray diffraction (XRD) patterns were collected by a Rigaku SmartLab diffractometer (Cu K*α*). The magnetic measurements were performed by an MPMS SQUID magnetometer from Quantum Design.

## III. RESULTS AND DISCUSSIONS

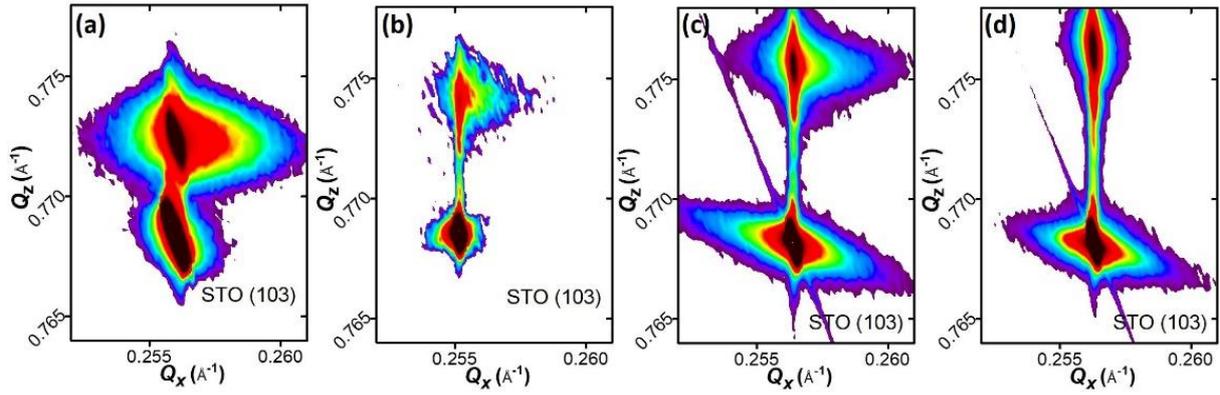

*Fig. 1 Reciprocal space maps of the LNMO films (a) L5, (b) L35, (c) L65, and (d) L95. The sample numbers represent the annealing time in minutes.*

The XRD and the XRR patterns of the LNMO films with different annealing times are shown in *supplementary* Fig. S1 (a) and (b), respectively. Reciprocal space maps of the films around the STO (103) reflection are shown in Fig. 1 (a) to (d). All the films are in-plane locked to the substrate. The out-of-plane lattice constant decreases as the annealing time is increased. The variation of the lattice constant as a function of annealing time is shown in Fig. 2 (a). The trend in the variation of the lattice constant with annealing time is reminiscent of the oxygen deficiency in perovskites.[15-17] The oxygen vacancies cause a lattice expansion and an increase in the lattice constant.[17] The oxygen vacancies formed in the as-deposited film (L5) are reduced by in-situ annealing in oxygen. This caused a reduction in the lattice constant for the films annealed at a longer duration and approached stoichiometry. The thickness of the films estimated from the XRR is ~70 nm for all four films.

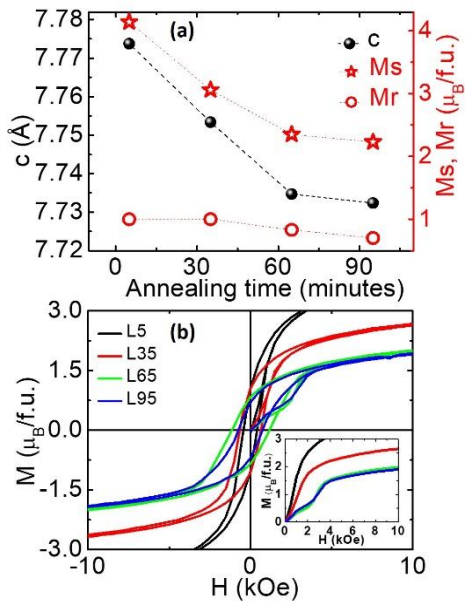

*Fig. 2 (a) Out-of-plane lattice constant, c, saturation magnetization, $M_S$, and remanent magnetization, $M_r$, of the LNMO thin films as a function of annealing time. (b) Enlarged view of the MH loop of LNMO films at 5 K. The inset is the virgin curve of the MH loop at 5 K.*

The substrate to film misfit ($f$) is calculated for the films as $f = (a_s-a_f)/a_f$, where, $a_s$ is the pseudocubic lattice constant of the substrate and $a_f$ is that of the film. By taking $a_f$ as the observed out-of-plane lattice constant ($c$), the misfit is re-calculated for each deposited film as shown in *supplementary* Fig. S2 (a). The misfit is increasing with the annealing time as the lattice constant is decreasing. The pseudocubic lattice constant analogous to bulk ($a_{bulk}$) perovskite is calculated from the in-plane ($\varepsilon_a$) and out-of-plane ($\varepsilon_c$) strain by taking the Poisson's ratio as $v = 0.3$, as described elsewhere.[15, 18, 19] The result is summarized in *supplementary* Fig. S2 (a). $a_{bulk}$ decreases with the annealing time and approaches the reported bulk value of 3.876 Å of the stoichiometric LNMO.[14, 20] The $\varepsilon_a$ is increasing with annealing time, as shown in *supplementary* Fig. S2 (b).

The magnetic hysteresis (*MH*) loop at 5 K for the LNMO films is shown in *supplementary* Fig. S3. Prior to each measurement, the sample temperature was raised to 350 K to clear memory of any previous magnetic ordering of the sample. The saturation magnetization ($M_S$) decreases as the annealing time is increased, as depicted in Fig. 2 (a). The $M_S$ is 4.15 $\mu_B$/f.u. for L5, which got reduced to 2.23 $\mu_B$/f.u. for L95. The reduced $M_S$ for the as-deposited sample compared to the theoretically expected value of 5 $\mu_B$/f.u. is due to the presence of oxygen deficiencies and anti-site defects.[21] Further annealing reduced the concentration of oxygen defects; at the same time, $M_S$ is found to be decreasing. This is due to the enhanced epitaxial strain and the formed anti-phase boundaries, as reported for LNMO films deposited on different substrates.[13, 14]

An enlarged view of the *MH* loop for the films is shown in Fig. 2 (b). The virgin curve lies outside the main loop for L65 and L95 with a metamagnetic-like behavior, as can be seen in the inset of Fig. 2 (b). Such behavior is an indication of the presence of antiferromagnetic ground states, as reported in other perovskites.[22, 23] This is also a manifestation of the antiferromagnetic anti-phase boundaries in the longer annealed films.[10, 22-24] The extra antiferromagnetic phase in the annealed samples subsequently reduced the total $M_S$.

For all the samples, the remanent magnetization ($M_r$) is less than or equal to 1 $\mu_B$/f.u., as shown in Fig. 2 (a). $M_r \gtrsim 1$ $\mu_B$/f.u. is reported to be an indication of off-stoichiometric samples.[12, 14] In the present work, $M_r$ is found to be 1 $\mu_B$/f.u. for L5 and L35, which is decreasing with extra annealing time as the oxygen deficiency is further reduced. However, the $M_r$ does not show a value above 1 $\mu_B$/f.u.. This confirms that the present films do not have any cation deficiency as reported before.[12]

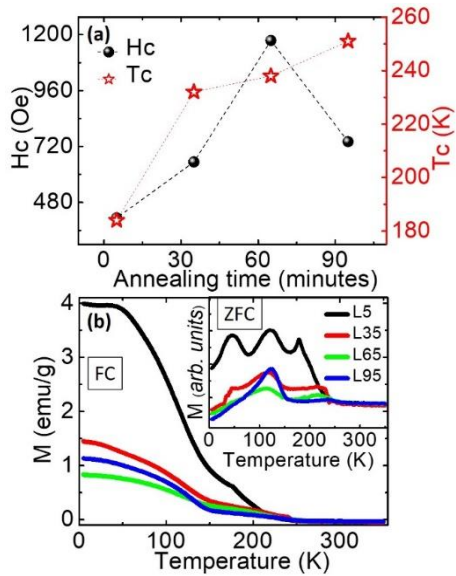

*Fig. 3 (a) $H_C$ and $T_C$ as a function of annealing time. $T_C$ is obtained from the minimum of the derivative plot dM/dT vs. T of FC M vs. T at 10 kOe. (b)The FC and ZFC (inset) magnetic moment of the LNMO films as a function of temperature for a range of 5-350 K with a magnetic field of 50 Oe.*

The coercivity ($H_C$) is 414 Oe for L5, which increased to 1175 Oe upon increasing the annealing time to 65 minutes for L65. On further increase to 95 minutes, the $H_C$ is reduced to 741 Oe. The variation of $H_C$ as a function of annealing time is shown in Fig. 3 (a). The $H_C$ of the high-$T_C$ ($T_{C1}$) phase is less than 320 Oe.[10] The higher coercivity indicates the existence of the low-$T_C$ ($T_{C2}$) phase in the films.[10] The anisotropy of the antiferromagnetic phase also prompted the enhanced coercivity in the LNMO films.

The magnetic moment was measured as a function of temperature (*M vs. T*) with different magnetic fields under field-cooled (FC) and zero-field cooled (ZFC) conditions. The FC and ZFC *M vs. T* at 50 Oe are shown in Fig. 3 (b) and Fig. 3 (b) inset, respectively. All the films possess the $T_{C1}$ and $T_{C2}$ phases. This is due to the biphasic nature of LNMO deposited at a considerably higher $pO_2$.[21] The ZFC *M vs. T* at 50 Oe for all the films shows peaks corresponding to the $T_{C1}$ and $T_{C2}$ phases. The peak around 45 K, which is due to the spin-glass nature of LNMO, is absent for the films L65 and L95.[3] Previous reports suggest the existence of two spin-glass transitions in LNMO, around 45 K and 110 K, originating from anti-site defects and oxygen off-stoichiometry.[3, 25] However, by controlling the oxygen defects, we were able to suppress the spin-glass anomaly at 45 K for the films L65 and L95.

The ZFC *M vs. T* at 100 Oe, 200 Oe, 500 Oe, and 800 Oe for the films L5, L35, L65, and L95 are shown in *supplementary* Fig. S4 (a) - (d). The peak at ~45 K is absent for L65 and L95 with an applied field of 100 Oe and 200 Oe. This peak appears with the application of 500 Oe and 800 Oe for these two films. For L5 and L35, the ~45 K spin-glass anomaly is visible even with a lower applied magnetic field due to the increased oxygen deficiency. The peak around ~110 K is present for all the films and is prominent for the longer annealed samples because of the anti-site defects present in the LNMO films.

The high-temperature $T_{C1}$ of the LNMO films (shown in Fig. 3 (a)) is obtained from the derivative of the FC *M vs. T* at 10 kOe. The $T_{C1}$ increases as the annealing time increases, which is due to the decreased oxygen defect formation in the lattice upon annealing. With the increased duration of post-annealing, the $T_{C1}$ of the LNMO films is increased from 184 K to 251 K. The oxygen annealing reduced the oxygen defect concentration and thereby enhanced the long-range ferromagnetic ordering. The superexchange interaction between the Ni and Mn *d*-orbitals is mediated via the oxygen *p*-orbitals. Annealing in oxygen reduced the oxygen deficiency in the films and gave a favorable way for the ferromagnetic superexchange interaction.

## IV. CONCLUSIONS

High-quality thin films of LNMO were fabricated by PLD at 900 °C and $pO_2$ of 500 mTorr. Oxygen defect concentration was formed during the growth of LNMO and was controlled and reduced by post-annealing treatments. The saturation magnetization is found to be 4.15 $\mu_B$/f.u. for the as-deposited samples, which decreased to 2.23 $\mu_B$/f.u. with post-annealing as a result of the epitaxial strain and anti-site defects. The magnetic transition temperature is 184 K for the as-deposited film. Upon increasing the annealing time, the transition temperature is increased to 251 K due to the low oxygen-defect concentration. This work highlights the importance of oxygen defect engineering to enhance the transition temperature of the ferromagnetic double perovskite LNMO.

## ACKNOWLEDGMENTS

The authors acknowledge the Deutsche Forschungsgemeinschaft (DFG) funding under project 429646908.

# Oxygen Defect Engineered Magnetism of La$_2$NiMnO$_6$ Thin Films

Jasnamol P. Palakkal, Thorsten Schneider, Lambert Alff

*Institute of Materials Science, Technische Universität Darmstadt, 64287 Darmstadt, Germany*

**Supplementary Information**

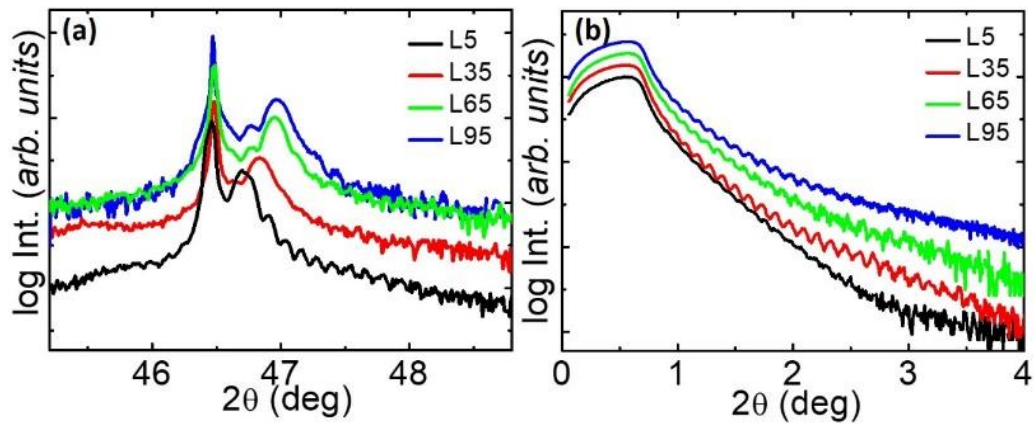

*Fig. S1 (a) The XRD pattern of LNMO films around the STO (002) reflection. (b) The XRR pattern of LNMO films. The sample numbers represent the annealing time in minutes.*

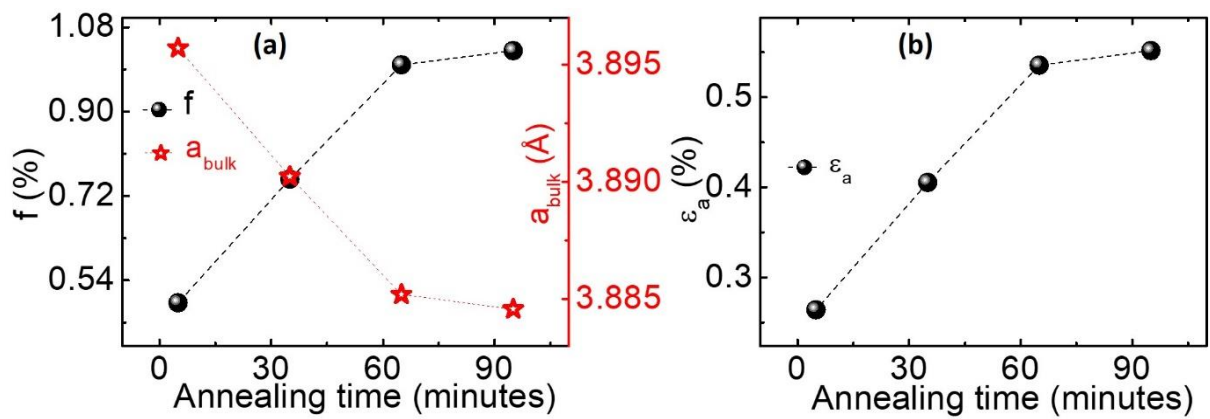

*Fig. S2 (a) Substrate to film misfit (f) and the calculated pseudocubic bulk lattice constant ($a_{bulk}$) of the LNMO films. (b) In-plane strain ($\varepsilon_a$) of the LNMO films.*

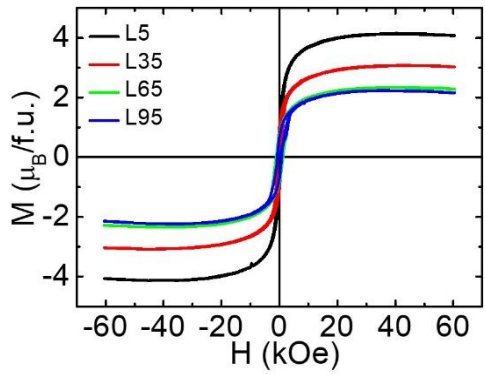

*Fig. S3 (b) The magnetic hysteresis loop of the LNMO films at 5 K.*

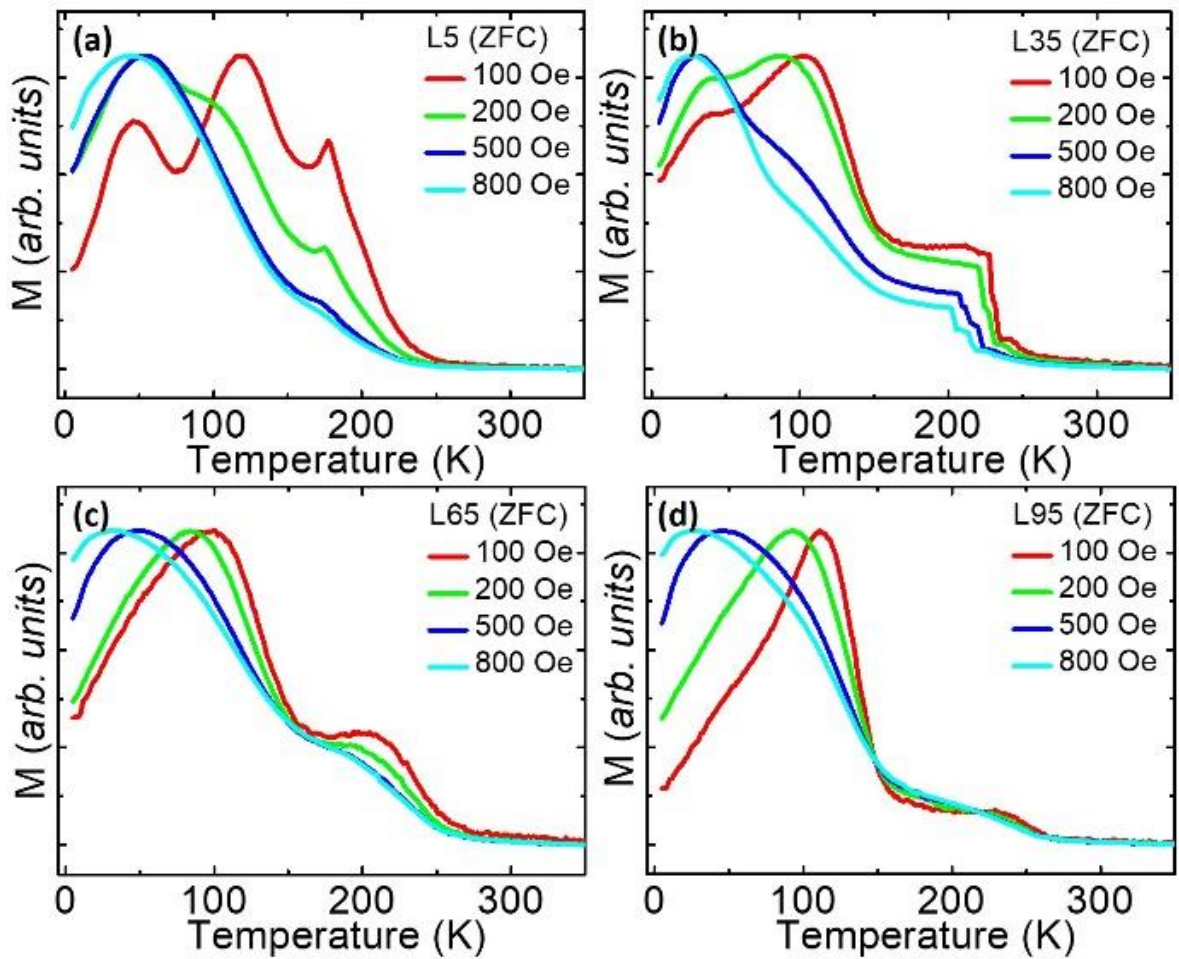

*Fig. S4 (a) The ZFC magnetic moment as a function of temperature at different magnetic fields for films (a) L5, (b) L35, (c) L65, and L95.*